\documentstyle{article}

\title{Superanalogs of symplectic and contact geometry and their
applications to quantum field theory.}
\author{Albert Schwarz\thanks {Research supported in part by NSF
grant No.
DMS-9201366}\\
Department of Mathematics, University of California,\\ Davis, CA
95616\\
ASSCHWARZ@UCDAVIS.EDU}
\date{}

\begin{document}
 \maketitle
 \smallskip
 \begin{abstract}
  The paper contains a short review of the theory of symplectic and
contact manifolds and of the generalization of this theory to the case of
supermanifolds. It is shown that this generalization can be used to
obtain some important results in quantum field theory. In particular,
regarding $N$-superconformal geometry as particular case of contact
complex geometry, one can better understand $N=2$ superconformal
field theory and its connection to topological conformal field theory.
The odd symplectic geometry constitutes a mathematical basis of
Batalin-Vilkovisky procedure of quantization of gauge theories.

    The exposition is based mostly on published papers. However, the
paper contains also a review of some unpublished results (in the
section devoted to the axiomatics of $N=2$ superconformal theory and
topological quantum field theory). The paper will be published in
Berezin memorial volume.

\end{abstract}

{\bf Introduction.}

   It is a great pleasure for me to publish my paper in this volume. F.
Berezin made extremely important contribution to mathematical
physics in many different directions. However, the most significant part
of his heritage is related to the idea that the theory of fermions
becomes very similar to the theory of bosons, if the usual functions ("
functions of commuting variables") are replaced by the "functions of
anticommuting variables" (elements of a Grassmann algebra). He was
the first who realized that along with standard algebra, analysis and
geometry one can construct algebra and analysis of functions
depending not only on commuting, but also on anticommuting
variables, and develop geometry of manifolds with commuting and
anticommuting coordinates. These ideas found very important
applications to physics. They were used to analyze a new kind of
symmetry. This symmetry, mixing bosons and fermions, is called
supersymmetry; therefore corresponding mathematical concepts are
also provided with the prefix "super".

   The present paper is based completely on such concepts. It begins
with a brief introduction to the ideas of supergeometry. The main part
of the paper contains a short review of the theory of symplectic and
contact manifolds and of the generalization of this theory to the case of
supermanifolds. It is shown that this generalization can be used to
obtain some important results in quantum field theory. In particular,
regarding $N$-superconformal geometry as a particular case of
contact complex geometry, one can better understand $N=2$
superconformal field theory and its connection to topological conformal
field theory. The odd symplectic geometry constitutes a mathematical
basis of the Batalin-Vilkovisky procedure of quantization of gauge
theories.

   Our exposition is based mostly on the papers [1]-[7]. However, the
paper contains also a review of some unpublished results (in the
section devoted to the axiomatics of $N=2$ superconformal theory and
topological quantum field theory).

   {\bf Supergeometry.}

   A smooth $m$-dimensional manifold can be defined as an object
obtained from domains in $R^m$ pasted together by means of smooth
transformations. This definition can be formulated in a purely algebraic
way. Namely, one can identify a domain $U\subset R^m$ with the
algebra $C^{\infty}(U)$ of all smooth functions on $U$ and a smooth
map of $U$ into $V$ with a homomorphism of  $C^{\infty}(V)$ into
$C^{\infty}(U)$. Such an algebraic construction can be generalized as
follows. By definition, we identify an $(m|n)$-dimensional superdomain
$U_n$ with the $Z_2$-graded algebra  $C^{\infty}(U)\otimes \Lambda
_n$ where $U$ is a domain in $R^m$ and   $\Lambda _n$ is a
Grassmann algebra with $n$ generators $\xi^1,...,\xi^n$. In particular if
$U=R^m$ we obtain a superdomain denoted by $R^{m|n}$. One says
that $R^{m|n}$ is an $(m|n)$-dimensional linear superspace. A map of
$U_n$ into $V_{n^{\prime}}$, where $U$ is a domain in $R^m,\  V$ is
a domain in $R^{m^{\prime}}$, is defined as an even homomorphism
of  $C^{\infty}(V)\otimes \Lambda _{n^{\prime}}$ into
$C^{\infty}(U)\otimes \Lambda _n$. (We say that an operator acting on
$Z_2$-graded spaces is even if it is parity preserving and odd if it is
parity reversing.) Elements of the algebra  $C^{\infty}(U)\otimes
\Lambda _n$ can be written as formal linear combinations
\begin {equation}
F=\sum_k\sum_{i_1,...,i_k} f_{i_1,...,i_k}(x)\xi^{i_1}...\xi^{i_k} ,
\end {equation}
where  $f_{i_1,...,i_k}(x)$ are smooth functions on $U$ and
$\xi^i\xi^j=-\xi^j\xi^i$. Without loss of generality we assume that the
coefficients $f_{i_1,...i_k}$ are antisymmetric with respect to a
permutation of $i_1,...,i_k$. It is convenient to consider the elements of
$C^{\infty}(U)\otimes \Lambda _n$ as functions depending on
commuting variables $(x^1,...,x^m)\in U$ and anticommuting variables
$\xi^1,...,\xi^n$. (In other words elements  $C^{\infty}(U)\otimes
\Lambda _n$ are considered as functions on the superdomain $U_n$
with $m$ commuting coordinates $x^1,...,x^m$ and $n$ anticommuting
coordinates $\xi^1,...,\xi^n$). A map of a superdomain $U_n$ with
coordinates $x^1,...,x^m,\xi^1,...,\xi^n$ into a superdomain
$V_{n^{\prime}}$ with coordinates  $\tilde {x}^1,...\tilde
{x}^{m^{\prime}},\tilde {\xi}^1,...,\tilde {\xi}^{n^{\prime}}$ can be
specified by the formulas
$$\tilde {x}^i=a^i(x^1,...x^m,\xi ^1,...,\xi^n)$$
$$\tilde {\xi}^j=\alpha ^j(x^1,...x^m,\xi ^1,...,\xi^n)$$
where $a^i,\  1\leq i\leq m^{\prime},$ and $\alpha ^j,\  1\leq j\leq
n^{\prime}$ are correspondingly even and odd elements of

$C^{\infty}(U)\otimes \Lambda _n$. (It is easy to check that the
substitution of $a^i$ and $\alpha ^j$  into the functions of variables
$\tilde {x}^i, \tilde {\xi}^j$ determines a homomorphism from
$C^{\infty}(V)\otimes \Lambda _{n^{\prime}}$ into $C^{\infty}(U)\otimes
\Lambda _n$ if the functions $a^i(x^1,...,x^m,0,...,0)$, $1\leq i\leq
m^{\prime}$ determine a map of the domain $U\subset  R^m$ into the
domain $V\subset R^{m^{\prime}}$.)

  Now one can define an $(m|n)$-dimensional supermanifold as an
object pasted together from $(m|n)$-dimensional superdomains by
means of invertible maps.
The definition of superdomain and supermanifold given above is
completely algebraic. In this definition a supermanifold "has no points."
However, if $M$ is a supermanifold and $\Lambda$ is an arbitrary
Grassmann algebra, one can construct the set $M_{\Lambda}$ of
$\Lambda$-points of $M$. In the particular case when $M$ is a
superdomain $U_n$ we define $M_{\Lambda}$ as the set of all rows
$(x^1,...,x^m,\xi^1,...,\xi^n)$ where $\xi^1,...,\xi^n$ are arbitrary odd
elements of $\Lambda$ and $x^1,...,x^m$ are even elements of
$\Lambda$ obeying $(m(x^1),...,m(x^m))\in U$ (here $m$ is the
standard homomorphism of $\Lambda$ onto $R$.) It is easy to check
that a map of superdomains generates a map of corresponding sets of
$\Lambda$-points. Using this remark one can define the set
$M_{\Lambda}$ for an arbitrary  supermanifold $M$. To every parity
preserving homomorphism $\rho :\Lambda\rightarrow
\Lambda^{\prime}$ of  Grassmann algebras one can assign a map
$\tilde {\rho}:M_{\Lambda}\rightarrow M_{\Lambda^{\prime}}$ in natural
way; if $\rho :\Lambda\rightarrow \Lambda ^{\prime}$and $\rho
^{\prime}:\Lambda^{\prime}\rightarrow \Lambda^{\prime \prime}$ are
two parity preserving homomorphism of  Grassmann algebras then $
\widetilde {\rho\rho^{\prime}}=\tilde{\rho}\cdot\tilde{\rho}^{\prime}$. In
other words a supermanifold $M$ determines a functor on the category
of  Grassmann algebras with values in the category of sets. (If $M$ is
an $(m|n)$-dimensional manifold and $\Lambda$ is a  Grassmann
algebra having $l$ generators, one can consider $M_{\Lambda}$ as a
smooth manifold of dimension $2^{l-1}(m+n)$. Therefore one can say
also that a  supermanifold determines a functor on the category of
Grassmann algebras with values in the category of smooth manifolds.)
The language of $\Lambda$-points is often very convenient.

  We define an algebraic operation on the supermanifold $M$ as a
map $M\times M\rightarrow M$. It is easy to see that such an
operation in $M$ determines naturally a map $M_{\Lambda}\times
M_{\Lambda}\rightarrow M_{\Lambda}$, i.e an operation on
$M_{\Lambda}$. (We are talking here about binary operations;
however one can consider in the same way operations with arbitrary
number of arguments.). Let us suppose that $M$ is equipped with a
binary operation and a unary operation in such a way that
$M_{\Lambda}$ is a Lie group with respect to corresponding
operations in $M_{\Lambda}$ for every $\Lambda$. (The operations
are considered as multiplication and taking the inverse respectively.)
We say then that $M$ is provided with a structure of Lie supergroup.
(In such a way a Lie supergroup determines a functor on the category
of  Grassmann algebras with values in the category of Lie groups.) The
notions of super Lie algebra, supercommutative algebra etc. can be
defined in similar way.

  We will define a superspace as a functor on the category of
Grassmann algebras taking values in the category of sets. (Morphisms
in  the category of Grassmann algebras are parity preserving
homomorphism.) Similarly, a supergroup can be defined as a functor
on  the category of Grassmann algebras taking values in  the category
of groups. One can define in natural way the notion of action of a
supergroup on a superspace and the notion of orbit  superspace of
such an action. It is clear that a supermanifold can be considered as a
superspace and a Lie supergroup can be regarded as a supergroup.
However if a Lie supergroup acts on a supermanifold the
corresponding orbit  superspace is not necessarily a supermanifold.

 Almost all notions of algebra, geometry and analysis can be
formulated for supermanifolds. For example, the (Berezin) integral of
the function (1) over $U_n$ is defined by the formula

\begin {equation}
\int_{U_n}Fd^nxd^n\xi=n!\int_U f_{1,...,n}(x)d^nx
\end {equation}
A (left) derivation in $C^{\infty}(U)\otimes \Lambda_n$ can be defined
as a linear operator $D$ satisfying the condition

$$D(uv)=Du\cdot v+(-1)^{\varepsilon (D)\varepsilon(u)}u\cdot Dv.$$
(Here $\varepsilon (u)$ denotes the parity of $u$ and $\varepsilon(D)$
is the parity of $D$.) The left  derivative ${\partial _l\over\partial \xi^i}$
with respect to an anticommuting variable $\xi^i$ is defined as the odd
derivation obeying ${\partial _l \over\partial \xi^i}\xi^j=\delta_i^j,\
{\partial_l\over \partial\xi^i}f(x)=0$. It is easy to check that  derivations
in $C^{\infty}(U)\otimes \Lambda_n$ can be identified with first order
differential operators (operators of the form $A^i(x,\xi){\partial \over
\partial x^i}+\alpha^j(x,\xi){\partial_l\over \partial \xi^j}$. The definitions
of right derivation and right derivative ${\partial_r\over \partial \xi^i}$
are similar. To define a differential form on a superdomain $U_n$ with
commuting coordinates $(x^1,...,x^m)\in U$ and anticommuting
coordinates $\xi ^1,...,\xi^n$ we consider a function of commuting
variables $x^1,...,x^m,\tilde {\xi}^1,...,\tilde {\xi}^n$ and anticommuting
variables $\xi ^1,...,\xi ^n,\tilde {x}^1,...\tilde {x}^m$. Such a function
determines a $k$-form if it is a polynomial of degree $k$ with respect
to the variables $\tilde {x}^1,...,\tilde {x}^m,\tilde {\xi}^1,...,\tilde
{\xi}^n$.
The variables $\tilde {x}^i, \tilde {\xi}^j$ can be identified with
differentials $dx^i, d\xi ^j$ (note that the parity of the differential is
opposite to the parity of corresponding  variable.) The generalization of
the Rham differential is defined by the formula

$$d=\sum _i\tilde {x}^i{\partial \over \partial x^i}+\sum _j
\tilde{\xi}^j{\partial _l \over\partial \xi ^j}.$$

   All notions defined above for superdomains are invariant with respect
to  invertible transformations and therefore our definitions work also for
supermanifolds. The only exception is the Berezin integral. As in the
case of the usual integral, the integrand acquires a factor equal to the
determinant of the Jacobian matrix by the change of variables. Of
course, the determinant here should be understood as the
superdeterminant (Berezinian).

   For a supermanifold $M$ one can introduce the notion of tangent
bundle $TM$ and cotangent bundle $T^*M$. One can consider also
the tangent  and cotangent bundles with the parity of fibers reversed;
we will use the notations $\Pi TM$ and $\Pi T^* M$ in this case. Note,
that a section of the  bundle $TM$ or of the bundle $\Pi TM$ can be
considered as a vector field on $M$; a section of $T^*M$ or of $\Pi T^*
M$ is a 1-form on $M$. A $k$-form on $M$ can be considered as a
function on $\Pi T M$.

   We will use freely the supergeneralizations of standard notions of
algebra and analysis.

  {\bf Symplectic supermanifolds.}

  We can apply the standard definition of symplectic manifold to the
case of  supermanifolds. Namely, a symplectic structure on a
supermanifold $M$ can be specified by means of an even
non-degenerate closed $2$-form

  \begin {equation}
  \omega =dz^a\omega_{ab}(z)dz^b.
  \end {equation}
  Here $z^a$ are (even and odd) coordinates on $M$. As always
$\omega$ is closed if $d\omega =0$ and non-degenerate if the (super)
matrix $\omega_{ab}$ is invertible. If the form $\omega$ in the
definition above is odd we say that it specifies an odd symplectic
structure; the   manifolds equipped with an odd symplectic structure
(odd symplectic manifolds or $P$-manifolds) will be considered at the
end of the paper. Symplectomorphisms (canonical transformations) of
a symplectic manifold $M$ are defined as transformations of $M$
preserving the form $\omega$ (i.e. $f$ is a symplectomorphism if
$f^*\omega =\omega$). Locally a $2$-form $\omega$ specifying a
symplectic structure in an appropriate coordinate system can be
written as

  \begin {equation}
  \omega _0=\sum_{1\leq i\leq n} dp_idx^i+\sum_{1\leq j\leq m}(d\xi
^j)^2
  \end {equation}
(Here $x^1,...,x^n,p_1,...,p_n$ are even coordinates, $\xi ^1,...,\xi^m$
are odd coordinates.) Therefore one can define a symplectic
supermanifold as a  supermanifold pasted together from superdomains
in $R^{2n|m}$ by means of transformations preserving $\omega_0$
(canonical transformations).

  One can define the Poisson bracket of two functions $F,G$ on the
symplectic manifold $M$ by the formula

   \begin {equation}
  \{ F,G\} ={\partial _r F\over\partial z^a}\omega^{ab}(z){\partial _l
G\over\partial z^b}
  \end {equation}
where $\omega ^{ab}$ stands for the matrix inverse to $\omega
_{ab}$.

  The space $C(M)$ of functions on $M$ can be considered as a
$Z_2$-graded Lie algebra (super Lie algebra) with respect to this
operation. We can assign to every function $H$ a vector field $K_H$
on $M$ by the formula
   \begin {equation}
  K_H^a=\omega ^{ab}(z){\partial _l H\over \partial z^b}.
  \end {equation}
  The corresponding first order differential operator $\hat {K}_H$ acts in
the following way:
  \begin {equation}
  \hat {K}_HF=\{ F,H\} .
  \end {equation}
 Vector fields obtained by means of this construction (Hamiltonian
vector fields) preserve the form $\omega$; in other words they can be
considered as infinitesimal symplectomorphisms. The map
$H\rightarrow K_H$ is a homomorphism of the (super)Lie algebra
$C(M)$ into the (super)Lie algebra of vector fields (of first order
differential operators).

 If we omit the requirement of non-degeneracy of the form $\omega$ in
the definition of symplectic (super)manifold we obtain the definition of
presymplectic manifold. If $\omega$ is degenerate there exist vectors
$\xi^l$ satisfying $\omega _{kl}(z)\xi^l=0$. Such vectors are called null
vectors of the form $\omega$. The set $W_z$ of null vectors at the
point $z\in M$ can be considered as linear subspace of the tangent
space $T_z$ to $M$ at the point $z\in M$.

 We want to construct a symplectic manifold $M^{\prime}$
corresponding to the presymplectic  manifold $M$. With this aim we
identify the point $z\in M$ with the point $z+dz$ if $dz$ is a null vector
of $\omega$ (i.e. $\omega_{kl}dz^l=0$). To be more rigorous, we will
suppose that $W_z$ is a linear superspace whose dimension does not
depend on $z\in M$; this dimension will be denoted by $(d_1|d_2)$. By
means of Frobenius' theorem for every point $z\in M$ we can
construct at least locally a $(d_1|d_2)$-dimensional manifold $R$
satisfying 1) $z\in R$, 2) the tangent space to $R$ at anarbitrary point
$u\in R$ coincides with $W_u$. (The subspaces $W_z$ determine
$(d_1|d_2)$-dimensional distribution  $W$ on the manifold $M$. This
distribution is integrable; this means that for every two vector fields
$\xi_1(z)\in W_z,\  \xi_2(z)\in W_z$ their commutator  also belongs to
$W_z$. By the Frobenius'theorem we can construct for every point
$z\in M$ an integral manifold $R$ of the distribution $W$ containing
the point $z$.) We will identify two points of $M$ belonging to the
same manifold $R$. Then for every point $z\in M$ we can find a
neighborhood $U$ in such a way that by this identification we obtain
from $U$ a symplectic manifold $U^{\prime}$ (the form
$\omega^{\prime}$ specifying the symplectic structure on $U^{\prime}$
can be determined by relation $\pi ^*\omega^{\prime}=\omega$ where
$\pi$ denotes the natural projection of $U$ onto $U^{\prime}$). In
global consideration we can meet topological troubles. (The global
behavior of an integral manifold can be very complicated. Therefore by
the identification of points belonging to the same integral manifold we
can obtain from $M$ a complicated topological space $M^{\prime}$.
The space is not necessarily a manifold; moreover one can't assert the
points of $M^{\prime}$ are closed sets.) However in physics we do not
usually meet these troubles.

  {\bf Contact manifolds.}

   We will say that a $1$-form $\alpha$  on the domain $U\subset
R^{2n+1}$ determines a precontact structure on $U$. If $\alpha
^{\prime}=F\alpha$, where $F$ denotes a non-vanishing function on
$U$, we say that the form $\alpha$ and $\alpha ^{\prime}$ determine
the same precontact structure.  If the $1$-form $\alpha$ is
non-degenerate (i.e. the $(2n+1$-form $\alpha\wedge (d\alpha )^n$
does not vanish anywhere) we say that $\alpha$ determines a contact
structure on $U$. For instance, the form

   \begin {equation}
   \alpha =dz+{1\over 2}(pdq-qdp),\ z\in R,\ p\in R^n,\  q\in R^n
   \end {equation}
determines a contact structure on $R^{2n+1}$. This example is
universal in some sense: locally every non-degenerate $1$-form takes
the form (8) in an appropriate coordinate sistem.

  A transformation $\varphi$ is called a contactomorphism if

   \begin {equation}
   \varphi ^*\alpha =G\alpha
   \end {equation}
where $G$ is  a non-vanishing function on $U$. In other words a
contactomorphism  can be defined as a map preserving (pre)contact
structure. If  a precontact structure is defined on the domain $U\in
R^{2n+1}$ by means of the $1$-form $\alpha =\alpha_i(z)dz^i$ we can
determine a  presymplectic structure on the domain $R^*\times
U\subset R^{2n+2}$ by means of the closed $2$-form $\omega
=d\lambda$ where

   \begin {equation}
   \lambda =dt+t\alpha =dt+t\alpha _i(z)dz^i
   \end {equation}
(here $R^*$ denotes the set of non-zero real numbers:$R^*=\{ t|t\in R,\
t\not= 0\}$). The  presymplectic structure on $R^*\times U$ is called a
symplectization of the precontact structure in $U$. It is easy to check
that the $2$-form

   \begin {equation}
   \omega =d\lambda=dt\wedge \alpha +td\alpha
   \end {equation}
 is non-degenerate if and only if the form $\alpha$ is non-degenerate.
In other words one can say that a contact structure in $U$ can be
defined as a  precontact structure giving a symplectic structure in
$R^*\times U$ after symplectization.

   Let us consider a contactomorphism $\varphi$ of the domain $U$.
We will construct a transformation $\tilde {\varphi}$ of $R^*\times U$
by the formula

   \begin {equation}
   \tilde {\varphi} (t,u)=(G^{-1}t,\varphi (u))
   \end {equation}
where $G$ is defined by (9). It is easy to check that $ \tilde {\varphi}$
is a symplectomorphism of $R^*\times U$; one can say that $ \tilde
{\varphi}$ is obtained from $\varphi$ by means of symplectization. The
transformation $ \tilde {\varphi}$ is homogeneous of degree $1$ with
respect to $t$: if $ \tilde {\varphi}(t,u)=(t^{\prime},u^{\prime})$ then $
\tilde {\varphi}(\lambda t,u)=(\lambda t^{\prime},u^{\prime})$. One can
verify that every symplectomorphism $ \tilde {\varphi}$ of $R^*\times
U$ satisfying this condition generates a contactomorphism $\varphi$ of
$U$ by the formula (12). This assertion permits us to give a description
of infinitesimal contactomorphisms. Let us suppose for simplicity that
$u=(z,p,q),\  z\in R,\ p\in R^n,\  q\in R^n$ and $\alpha$ is given by (8).
It is evident that infinitesimal contactomorphisms correspond to
infinitesimal  symplectomorphisms  (to Hamiltonian vector fields) in
$R^*\times U$ having a function of the form

   \begin {equation}
   H(t,u)=H(t,z,p,q)=tK(z,p,q)
   \end {equation}
  as a Hamiltonian. We see that an infinitesimal contactomorphism can
be written in the form

   \begin {equation}
   \delta z=K-{p\over 2}{\partial K\over \partial p}-{q\over 2}{\partial
K\over \partial q},\ \delta p=-{\partial K\over \partial q}+{p\over
2}{\partial K\over \partial z},\  \delta q={\partial K\over \partial p}+{q\over
2}{\partial K\over \partial z}
   \end {equation}
  where $K$ is an arbitrary function on $U$.

  Let us define a $(2n+1)$-dimensional contact manifold as a manifold
pasted together from domains in $R^{2n+1}$ provided with contact
structure, by means of a contactomorphism. Without loss of generality
one can assume that the contact structure on these domains is
standard (determined by the form (8)). Given an arbitrary
$n$-dimensional manifold $M$, one can construct a
$(2n-1)$-dimensional contact manifold $PT^*M$ in the following way.
Let us consider the space $T^*M$ of all covectors in $M$ and the form
$\omega =dp_i\wedge dq^i$ specifying a symplectic structure on
$T^*M$ (here $q^1,...,q^n$ are local coordinates on $M$ and
$p_1,...,p_n$ denote coordinates of the covector). Let us identify the
points $(p_1,...p_n,q^1,...,q^n)$ and  $(\lambda p_1,...\lambda
p_n,q^1,...,q^n)$ in $T^*M\setminus M$ (here $\lambda \not= 0$). The
space obtained by means of this identification will be denoted by
$PT^*M$. (One can say that $PT^*M$ is the space of the projectivized
cotangent bundle.) The space $PT^*M$ can be provided with a contact
structure in natural way; the symplectization of this contact structure
gives the natural symplectic structure on $T^*M$. The contact
structure in $PT^*M$ can be specified by means of a $1$-form
$Fp_jdq^j$. More precisely, let us consider the case when $M$ is a
domain in $R^n$ and $q^1,...,q^n$ are coordinates on $M$. Then
$T^*M\setminus M$ can be covered by sets $\tilde {U}_1,...,\tilde
{U}_n$, where $\tilde {U}_i$ is singled out by the condition $p_i\not=
0$. By identification we obtain from  $\tilde {U}_1,...,\tilde {U}_n$ the
charts  $U_1,...,U_n$ covering $PT^*M$. We can introduce
coordinates in $U_i$ by the formula $\pi _k=p_k/p_i$. In other words
each point of $U_i$ can be obtained from a point of  $\tilde {U}_i$
satisfying $p_i=1$; the  coordinates of this point of  $\tilde {U}_i$ are
considered as coordinates of the corresponding point in $U_i$. The
$1$-form specifying the contact structure in $U_i$ can be written as

  \begin {equation}
  \alpha ^{(i)}=dq^i+\sum_{k\not= i}\pi_kdq^k=p_rdq^r.
  \end {equation}
  It is easy to check that the forms $\alpha ^{(i)}$ specify a contact
structure in $PT^*M$ (i.e. $\alpha ^{(i)}$ and $\alpha ^{(j)}$ are
proportional in $U_i\cap U_j$).  The definition of precontact structure
and the definition of contactomorphism can be applied also in the case
when the domain $U\subset R^{2n+1}$ is replaced by a superdomain
$U\subset R^{2n+1|m}$. The symplectization of precontact structure
can be defined in the supercase and gives a presymplectic structure
on the superdomain $R^*\times U\subset R^{2n+2|m}$. As usual, we
say that the $1$-form $\alpha$ determines a contact structure, if
$\alpha$ is non-degenerate, but the definition of non-degeneracy
should be changed. One can say, for example, that $\alpha$ is
non-degenerate if the $2$-form (11) is non-degenerate. (In other words
a contact structure in $U$ can be defined as a precontact structure
giving a symplectic structure in $R^*\times U$ after symplectization.)

  {\bf Symplectic and contact complex (super)manifolds.}

   Let $M$ be a complex (super)manifold. A presymplectic structure on
$M$ can be specified by a closed holomorphic $(2,0)$-form $\omega$
on $M$; if this form is non-degenerate we say that $M$ is a symplectic
complex manifold. In local coordinates $(z^1,...,z^n)$ the form
$\omega$ can be represented by the formula (3) where $\omega
_{ab}(z)$ are holomorphic functions.

    In such a way the definition of symplectic structure on a complex
manifold repeats the definition in the real case, but $\omega$ has to
be considered as a holomorphic $(2,0)$-form. The definition of contact
structure on a complex manifold also repeats the definition of contact
structure on a real manifold (the form $\alpha$ in this definition must
be considered as a non-degenerate holomorphic $(1,0)$-form).

    If $M$ is anarbitrary real or complex $(m|n)$-dimensional
supermanifold then the cotangent space $T^*M$ can be considered as
a $(2m|2n)$-dimensional symplectic supermanifold and the
projectivized cotangent space $PT^*M$  as a $(2m-1|2n)$-dimensional
contact supermanifold. (The construction of symplectic structure on
$T^*M$ and of contact structure in $PT^*M$ is quite similar to the
construction described above for the case when $M$ is a real
$m$-dimensional manifold).

    Complex contact geometry is closely related to superconformal
geometry. Recall that an $N$-superconformal transformation of a
$(1|N)$-dimensional complex  superdomain $U$ can be defined as a
complex analytic transformation preserving up to multiplier the
$1$-form

    \begin {equation}
   \alpha =dz+\sum _i\theta _id\theta _i
   \end {equation}

   Here $(z_1,\theta_1,...,\theta_n)$ denote complex coordinates in
$U$ ($z$ is even, $\theta_1,...,\theta_n$ are odd). The $1$-form
$\alpha$ is non-degenerate and therefore specifies complex contact
structure. Superconformal transformations can be interpreted as
complex contact transformations of the contact manifold $U$. An
$N$-superconformal manifold can be defined as a manifold pasted
together from $(1|N)$-dimensional complex superdomains by means
of  $N$-superconformal transformations. We see that an
$N$-superconformal manifold can be considered as a
$(1|N)$-dimensional complex contact manifold. Conversely, every
$(1|N)$-dimensional complex contact manifold can be considered as
an $N$-superconformal manifold. The proof is based on the fact that
locally every non-degenerate holomorphic form $\alpha
_0(z,\theta)dz+\sum \alpha_i(z,\theta )d\theta_i $ on a
$(1|N)$-dimensional superdomain in appropriate complex coordinate
system can be identified (up to holomorphic multiplier) with the form
(16). We see that a $(1|N)$-dimensional complex contact manifold can
be covered with local coordinates $(z,\theta _1,...,\theta _N)$ in such a
way that transition functions are  $N$-superconformal transformations.
(These coordinates are called superconformal coordinates.) Using
these coordinates one can introduce covariant derivatives

   $$D_i={\partial \over \partial \theta _i}+\theta_i{\partial\over\partial
z}.$$
   It is easy to check that covariant derivatives in superconformal
coordinate systems $(z,\theta _1,...,\theta _N)$ and  $(\tilde {z},\tilde
{\theta} _1,...,\tilde {\theta} _N)$ are connected by the relation

    \begin {equation}
   \tilde {D}_i=\sum_jF_{ij}(z,\theta )D_j
   \end {equation}
 (This fact follows immediately from the remark that the orthogonal
complement to the covector corresponding to the $1$-form (16) is
spanned by the vectors $D_1,...,D_N$.) In the case $N=2$ it is
convenient to introduce coordinates

$$\theta_+={1\over \sqrt {2}}(\theta_1+i\theta _2),\ \  \theta_-={1\over
\sqrt{2}}(\theta_1-i\theta_2)$$

and covariant derivatives

    \begin {equation}
   D_+={1\over \sqrt{2}}(D_1+iD_2)={\partial \over
\partial\theta_-}+{1\over 2}\theta_+{\partial\over \partial z},\
D_-={1\over \sqrt{2}}(D_1-iD_2)={\partial \over \partial\theta_+}+{1\over
2}\theta_-{\partial\over \partial z}
   \end {equation}
   One can verify that the behavior of $D_+,\  D_-$ under an $N=2$
superconformal transformation is given by the formulas

    \begin {equation}
   \tilde {D}_+=F_+D_+,\ \ \tilde {D}_-=F_-D_-
   \end {equation}
   (untwisted case) or by the formulas

   \begin {equation}
   \tilde {D}_+=G_+D_-,\ \  \tilde {D}_-=G_-D_+
   \end {equation}
   (twisted case). An $N=2$  superconformal manifold $M$ is called
untwisted if a local coordinate system in $M$ can be chosen in such a
way that the transition functions are untwisted  superconformal
transformations (covariant derivatives in different patches are related
by (19)). Let us denote by $M^{k|l}$ the moduli space of
   $(k|l)$-dimensional complex supermanifolds, i.e. the space of
classes of such manifolds with respect to holomorphic equivalence.
This definition is not rigorous. More precisely one can also define the
moduli space using as a starting point the superspace of all complex
structures on a fixed real supermanifold and factorizing this
superspace with respect to an appropriate equivalence relation. The
moduli space defined in such a way  is not necessarily a
supermanifold. However it can be considered as a superspace.

   The notion of moduli space of complex manifolds is closely related to
the notion of family of complex manifolds. In some sense the moduli
space can be considered as a base of a universal family. Recall that a
holomorphic family of compact complex $(k|l)$-dimensional
supermanifolds can be defined as holomorphic map of a complex
supermanifold $E$ onto a complex supermanifold $S$ having
$(k|l)$-dimensional compact complex manifolds as fibers. One says
that $S$ is the base of the family or that the family is parametrized by
the points of $S$. The definition of a continuous family is analogous.
(Speaking about families of manifolds we have in mind continuous
families.) For every continuous map $\rho :S^{\prime}\rightarrow S$ of
a space $S^{\prime}$ into the base $S$ of a family $F$ one can
construct in natural way a family $F^{\prime}$ with the space
$S^{\prime}$ as a base (the pullback of $F$).

   In similar way we can define other moduli spaces, for example the
moduli space $M_N$ of $N$-superconformal complex manifolds. (This
moduli space is related to families of $N$-superconformal manifolds.)

    For every  $(1|n)$-dimensional complex supermanifold $M$ we can
construct a  $(1|2n)$-dimensional complex contact supermanifold
$PT^*M$. Considering $PT^*M$ as $2n$-superconformal manifold we
obtain an embedding of the moduli space $M^{1|n}$ into the moduli
space $M_{2n}$. Let us consider more carefully the case $n=1$. In
this case we see that to every $(1|1)$-dimension complex
supermanifold $M$ we can assign the $N=2$  superconformal
manifold $\hat {M}=PT^*M$. If $(z,\theta)$ are complex coordinates in
$M$ then the points of $T^*M$ can be specified by coordinates
$(p,\rho ,z,\theta )$ where $(p,\rho)$ are the coordinates of the
covector on $M$. The behavior of these coordinates under the
transformation $(z,\theta )\rightarrow (\tilde {z},\tilde {\theta})$ is given
by

    \begin {equation}
   \tilde {p}={\partial z\over \partial\tilde {z}}p+{\partial z\over
 \partial\tilde
{\theta}}\rho,
   \end {equation}
   .
    \begin {equation}
  \tilde {\rho}={\partial \theta\over \partial\tilde {z}}p+{\partial
 \theta\over
\partial\tilde {\theta}}\rho.
   \end {equation}
A point of $PT^*M$ can be specified by a triple $(\rho,z,\theta)$ and
the behavior of this triple by the transformation
 $(z,\theta)\rightarrow
(\tilde {z},\tilde {\theta})$ is given by

    \begin {equation}
   \tilde {z}=\tilde{z}(z,\theta ),\ \
   \tilde {\theta}=\tilde {\theta}(z,\theta),\ \
   \tilde {\rho}=({\partial \theta \over \partial z}+{\partial
\theta\over
\partial \tilde {\theta}}\rho )({\partial z\over \partial \tilde {z}}+
{\partial
z\over \partial  \tilde {\theta}}\rho)^{-1}
   \end {equation}
   Using the general rule one can write the $1$-form specifying the
contact ($N=2$ superconformal) structure in $PT^*M$ as

    \begin {equation}
   \alpha =dz+\rho d\theta
   \end {equation}

   (we consider the form $pdz+\rho d\theta$ for $p=1$).

     In appropriate coordinates this form coincides with (16). Using
these coordinates one can check that the transformation (23) is an
untwisted $N=2$ superconformal transformation and therefore
$PT^*M$ is an untwisted $N=2$   superconformal  manifold. It is easy
to check that every untwisted $N=2$  superconformal  manifold can be
obtained by means of this construction (see ref. [2]). The proof can be
based on the remark that for every untwisted $N=2$ superconformal
manifold $N$ one can construct a $(1|1)$-dimensional complex
manifold $\alpha (N)$ factorizing $N$ with respect to the vector field
$D_+$. (In other wcrds, we assume that there exists a natural map $\pi
:N\rightarrow \alpha (N)$ and that the set of functions on $N$ having
the form $\Phi (\pi(x))$, where $\Phi$ is a function on $\alpha (N)$,
consists of functions annihilated by the operator $D_+$). One can
check that $P T^*\alpha(N)$ is isomorphic to $N$. In such a way we
have proved that there is one-to-one correspondence between classes
of $(1|1)$-dimensional complex manifolds and classes of $N=2$
untwisted   superconformal manifolds. In other words, the moduli
space $M^{1|1}$ coincides with the part of $M_2$ corresponding to
untwisted $N=2$  superconformal manifolds; we denote this part by
$\tilde {M}_2$. Note that all moduli spaces at hand have natural
complex structures; the isomorphism $\alpha$ is compatible with the
complex structures in $M_2$ and $M^{1|1}$. One can replace $D_+$
by $D_-$ in the construction of the isomorphism $\alpha :\tilde
{M}_2\rightarrow M^{1|1}$, we obtain another isomorphism $\beta
:\tilde {M}_2\rightarrow M^{1|1}$. This means that there exists a natural
involution $\iota :M^{1|1}\rightarrow M^{1|1}$ defined by the formula
$\beta =\iota \alpha$. This involution can be described geometrically in
the following way: if $M$ is a $(1|1)$-dimensional complex manifold,
we define $\iota (M)$ as the manifold consisting of all
$(0|1)$-dimensional submanifolds of $M$.

     The superMumford form in superstring theory can be considered as
an object defined on $M^{1|1}$; using this fact and the connection
between $M^{1|1}$ and $M_2$ one can discover hidden $N=2$
superconformal symmetry of the superMumford form [2]. A similar
consideration leads to the explanation of hidden $N=2$
superconformal symmetry of $B-C$ system discovered in [8].

  {\bf Axiomatics of $N=2$ superconformal theory and topological
quantum field theory .}

 Let us recall briefly Segal's axiomatics of conformal field theory (CFT).

 Let us consider a compact complex one-dimensional manifold $M$
with $m+n$ embedded non-overlapping parametrized disks
$D_1,...,D_m,D_1^{\prime},...,D_n^{\prime}$. (The set of disks is
divided in two parts: "incoming" disks $D_1,...,D_m$ and "outgoing"
$D_1^{\prime},...,D_n^{\prime}$.) The moduli space of such objects will
be denoted by $P_{m,n}$. There exist natural maps

 \begin {equation}
 P_{m_1,n_1}\times P_{m_2,n_2}\rightarrow P_{m_1+m_2,n_1+n_2},
 \end {equation}
 \begin {equation}
 P_{m,n}\rightarrow P_{m-1,n-1}.
 \end {equation}
 (To construct the first map we take the disjoint union of manifolds. The
second map corresponds to sewing of $D_m$ with $D_n^{\prime}$. In
other words we identify the points $P\in D_m$ and $P^{\prime}\in
D_m^{\prime}$ if corresponding coordinates are related by the formula
$z\cdot z^{\prime}=1/4$ and delete the points with $|z|<{1\over 2},\
|z^{\prime}|<{1\over 2}$. We assume that all disks are parametrized by
the points of the unit disk $D\subset C$.) In Segal's axiomatics we
should fix a linear space $H$. The main object is a linear map $H^m
\rightarrow H^n$ assigned to every point $M\in P_{m,n}$; this map
must have a trace. In other words we should have a map
$\alpha_{m,n}: P_{m,n}\rightarrow B_{m,n}=Map (H^m,H^n)$ of
$P_{m,n}$ into the space $B_{m,n}$ of trace class linear maps of
$H^m$ into $H^n$. (Here $H^m$ stands for the tensor product of $m$
copies of $H$). Axioms are the conditions of compatibility of maps
$\alpha_{m,n}$ and maps (25), (26). These axioms can be formulated
in terms of commutativity of diagrams containing the maps
$\alpha_{m,n}$, the maps (25), (26) and the natural maps

 \begin {equation}
 B_{m_1,n_1}\otimes B_{m_2,n_2}\rightarrow
B_{m_1+m_2,n_1+n_2},
 \end {equation}

 \begin {equation}
 B_{m,n}\rightarrow B_{m-1,n-1}.
 \end {equation}
One should require also the compatibility of the maps $\alpha _{m,n}$
with the natural action of $S_m\times S_n$ on $P_{m,n}$ and
$B_{m,n}$. (Here $S_m$ denotes the symmetric group.) The maps
$\alpha_{1|1}$ can be used to define the representation of Virasoro
algebra on $H$; more precisely we obtain two copies of the Virasoro
algebra with generators $L_n,\bar {L}_n$ obeying $[L_n,\bar
{L}_m]=0$. (The axioms above describe CFT with central charge
$c=0$. The maps $\alpha_{1|1}$ give in this case representations of
the Lie  algebra $diff(S^1)$ of vector fields on the circle. To describe
CFT with non-vanishing central charge one should assume that the
maps $\alpha_{m,n}$ are defined only up to a multiplier. Then the
maps $\alpha_{1|1}$ give rise to  projective representations  of
$diff(S^1)$, i.e. to  representations of the Virasoro algebra, the central
extension of $diff (S^1)$.)

It is important to emphasize that one can reformulate the Segal's
axiomatics using families of one-dimensional complex manifolds with
embedded disks instead of moduli spaces of such manifolds. We will
use the term $(m,n)$-family for a family of one-dimensional compact
complex manifolds with $m$ embedded "incoming" disks and $n$
embedded "outgoing" disks. The base of an  $(m,n)$-family $F$ will be
denoted by $B_F$. One should consider as a basic object a map
$\alpha _{m,n}^F :B_F\rightarrow B_{m,n} =Map(H^m,H^n)$ defined
for every  $(m,n)$-family $F$. Repeating the definitions of the maps
(25), (26) we can obtain corresponding maps for families. (For
example, the construction of the map (26) permits us to assign an
$(m-1,n-1)$-family to every  $(m,n)$-family.) It is easy to translate the
axioms above into the language of families. One should add the
following axiom:

If the  $(m,n)$-family $F^{\prime}$ can be obtained as a pullback of an
$(m,n)$-family $F$ by the map $\rho :B^{\prime} \rightarrow B_F$,
then $\alpha_{m,n}^{F^{\prime}}=\alpha_{m,n}^F \rho$.

 Similar axiomatics for superconformal theory can be obtained if we
replace  one-dimensional complex manifolds with $N$-superconformal
manifolds. (The space $H$ should be $Z_2$-graded in this case.) It
follows from the consideration above that $N=2$ superconformal
theory has also another description: one should replace
one-dimensional complex manifolds with $(1|1)$-dimensional complex
manifolds and disks with superdisks. The simplicity of this description
leads to a conclusion that probably $N=2$ superconformal symmetry is
more fundamental than $N=1$ superconformal symmetry.

 To give an axiomatic description of topological conformal field theory
we introduce a notion of a ${\cal T}$-manifold. One can define a ${\cal
T}$-manifold as a $(1|1)$-dimensional complex manifold equipped with
a non-degenerate even exact $1$-form $\omega$ (non-degeneracy
means here that  $\omega$ can be represented in the form $\omega
=d\varphi$ where locally the odd function  $\varphi$ can be written as
$\alpha (z)+a(z)\theta$ with  invertible $a(z)$.). The notion of a ${\cal
T}$-manifold is closely related to the notion of semirigid super
Riemann surface [9]. We define topological conformal field theory
(TCFT) replacing complex manifolds with  ${\cal T}$-manifolds in
Segal's axiomatics. Following the construction of operators $L_n,\bar
{L}_n$ in CFT we can construct operators in $Z_2$-graded space $H$.
We obtain even operators $L_n,\bar {L}_n$, generating two commuting
copies of the Virasoro algebra, and odd operators $b_n, \bar {b}_n, Q,
\bar {Q}$ obeying $L_n=[b_n,Q]_+,\  \bar {L}_n=[\bar{b}_n,\bar {Q}]_+$
(all other anticommutators of odd operators vanish). For every $N=2$
superconformal theory we can construct a TCFT in an obvious way
utilizing the fact that there is a natural map $\tau$ of the moduli space
$M_{\cal {T}}$ of  ${\cal T}$-manifolds into $M^{1|1}$ and that this map
can be extended to a map of the moduli space of manifolds with
embedded disks. (Every  ${\cal T}$-manifold can be considered as
$(1|1)$-dimensional complex manifold.)  Moreover, we can construct
two different TCFT corresponding to a given $N=2$ superconformal
theory using the involution in $M^{1|1}$ decribed above (we use the
map of $M_{\cal {T} }$ into $M^{1|1}$ defined as a composition of the
map $\tau :M_{\cal {T}}\rightarrow M^{1|1}$ with this involution). The
construction above can be an  considered as a geometric counterpart
of Witten's twisting. The two TCFT corresponding to $N=2$
superconformal theory are known as the $A$-model and $B$-model.

  Note that in the definition of TCFT it is convenient to work with the
version of Segal's axiomatics based on the consideration of families of
manifolds. (The moduli space of $\cal {T}$-manifolds is not a
supermanifold.) If the map $p: E\rightarrow B$ determines a
holomorphic family $F$ of one-dimensional complex manifolds, then
one can construct a family $\tilde {F}$ of $\cal {T}$-manifolds in the
following way. Let us consider a supermanifold $\tilde {B}=\Pi TB$ (the
space of the tangent boundle with reversed parity) and the natural map
$\Phi : \Pi TB\times C^{0|1}\rightarrow B$. (In local coordinates $\Phi $
transforms the point $(m,\mu,\nu )$ where $m\in B$, $\mu$ is a
tangent vector at the point $m$, and $\nu \in C^{0|1}$ into $m+\mu
\nu$.). Then we can define $\tilde {E}$ as a space consisting of points
$(e,m,\mu ,\nu )\in E\times \tilde {B}\times C^{0|1}$ obeying $p(e)=\Phi
(m,\mu, \nu )$. The maps $\tilde {p}: \tilde {E}\rightarrow \tilde {B}$ and
$\pi :\tilde {E}\rightarrow C^{0|1}$ are defined by means of projections
of the direct product $E\times \tilde {B}\times C^{0|1}$ onto the factors
$\tilde {B}$ and $C^{0|1}$. The map $\tilde {p}:\tilde {E}\rightarrow
\tilde {B}$ determines a family of $(1|1)$-dimensional complex
manifolds; the map $\pi$ determines a structure of $\cal {T}$-manifold
on the fibers of $\tilde {p}$. We obtained therefore holomorphic family
$\tilde {F}$ of $\cal {T}$-manifolds. Applying this construction to the
family corresponding to the moduli space $P_{m,n}$ we obtain a family
of $\cal {T}$-manifolds with embedded superdisks; the base of this
family is $\tilde {P}_{m,n} = \Pi TP_{m,n}$. The functions on $\tilde {P}
_{m,n}$ are forms on $P_{m,n}$; this fact permits us to relate the
axiomatics based on $\cal {T}$-manifolds with axiomatics of TCFT
formulated in [10].

  The approach to TCFT in [11] was based on the consideration of
two-dimensional manifolds equipped with metric tensor $g_{\alpha
\beta}$ and odd field $\varphi _{\alpha \beta}$. The  field $\varphi
_{\alpha \beta}$ can be considered as infinitesimal odd variation of the
metric tensor $g_{\alpha \beta}$; in other words the space $\tilde {Q}$
of pairs $(g_{\alpha \beta}, \varphi _{\alpha \beta})$ can be
represented as $\Pi TQ$ where $Q $ is the space of metrics on a
given two-dimensional manifold. After factorization with respect to the
group of diffeomorphisms and Weyl group we obtain the moduli space
$M$ of complex structures. Similarly, after appropriate factorization we
obtain $\Pi TM$ from $\tilde {Q}=\Pi TQ$. Using this remark one can
construct a $\cal {T}$ -manifold for every two-dimensional manifold
equipped with tensors $g_{\alpha \beta}, \varphi _{\alpha \beta}$.

  {\bf Odd symplectic geometry and BV-formalism.}

  We mentioned already that odd symplectic structure on a
supermanifold $M$ is determined by a non-degenerate closed odd
2-form $\omega$. Odd symplectic manifolds($P$-manifolds) play an
important role in the Batalin-Vilkovisky quantization procedure; we will
formulate here some results about the notion of $P$-manifold and
related notions.

  One can check that a $P$-manifold can be covered by local
coordinate systems in such a way that in every chart the form
$\omega$ is represented as $dx^id\xi_i$ (here $x^1,...,x^n $ are even
and $\xi_1,...,\xi_n$ are odd coordinates). In other words a
$P$-manifold can be pasted together from superdomains by means of
transformations preserving the form $dx^id\xi_i$. It is important to
emphasize that the transformations preserving the form $dx^id\xi_i$
are not necessarily volume  preserving. (By definition, a
transformation of a superdomain is volume preserving if the
determinant of Jacobian matrix is equal to $1$. Of course we are
talking here about the  superdeterminant, or Berezinian). It is well
known that there exists a natural volume element in an even
symplectic manifold and symplectic  transformations are volume
preserving; we see that in the case of $P$-manifolds the situation is
different. We define an $SP$-manifold as a manifold $M$ pasted
together from $(n|n)$-dimensional superdomains by means of
transformations preserving the form $dx^id\xi_i$ and the volume. An
operator $\Delta$ acting on functions on an $SP$-manifold can be
defined by the formula

  \begin {equation}
  \Delta ={\partial^2\over \partial x^i\partial\xi _i}.
  \end {equation}

   It is easy to check that this formula gives a globally defined operator
obeying $\Delta^2=0$. Conversely, let

\begin {equation}
   \Delta =\omega ^{ij}(z){\partial ^2\over \partial z^i\partial
z^j}+\alpha^i(z){\partial\over\partial z^i}+\beta(z)
\end {equation}

 be an odd second order differential operator on supermanifold $M$
satisfying $\Delta^2=0$. One can prove that in the case when the
matrix $\omega ^{ij}(z)$ is non-degenerate the operator $\Delta$ can
be represented locally in the form (29). This means that such an
operator determines an $SP$-structure in the manifold $M$. In
particular, it determines a $P$-structure and a volume element on
$M$.

  Let us say that a submanifold $L$ of a $P$-manifold $M$ is
Lagrangian if in a neighborhood of every point of $L$ one can
introduce such a coordinate system $x^1,...,x^n,\xi_1,...,\xi_n$ such,
that $\omega$ has the standard form $dx^id\xi_i$ and $L$ is singled
out by the equations $\xi_1=0,...,\xi_n=0$. (Here $x^i$ can be even or
odd, the parity of $\xi_i$ is opposite to the parity of $x^i$. Therefore
dim$L=(k|n-k)$ where $0\leq k\leq n$.) In a more invariant way one
can characterize Lagrangian manifold as a maximal  submanifold  of
$M$ where the form $\omega$ vanishes. If $M$ is an $SP$-manifold
one can introduce a volume element in a  Lagrangian submanifold
$L\subset M$. For example one can use a local coordinate system
$x^1,...,x^n,\xi_1,...,\xi_n$ where $\Delta $ is standard and $L$ is
singled out by equations $\xi_1=...=\xi_n=0$; the volume element
$dx^1...dx^n$ in $L$ does not depend on the choice of coordinates up
to a sign. (One can prove the existence of the coordinate system that
we used.) In Batalin-Vilkovisky formalism physical quantities are
represented by integrals of $\exp ({1\over \hbar}S)$ over  Lagrangian
submanifolds. Here $S$ is the extended classical action that should
satisfy the so called quantum master equation $\Delta \exp ({1\over
\hbar}S)=0$. The following theorem constitutes a mathematical basis
of Batalin-Vilkovisky formalism.

 {\it Let $H,H^{\prime}, K$ be functions on compact $SP$-manifold
$M$ obeying $\Delta H=\Delta H^{\prime}=0,\  H^{\prime}-H=\Delta K$.
Then for every two compact homologous Lagrangian submanifolds
$L\subset M,\  L^{\prime}\subset  M$ we have }
\begin {equation}
   \int _L Hd\lambda=\int_{L^{\prime}} H^{\prime}d\lambda^{\prime}
\end {equation}

   Here $d\lambda$ and $d\lambda^{\prime}$ denote volume elements
induced by the $SP$-structure in $M$. Compactness and homology
are defined in terms of bodies of supermanifolds. $H$ and
$H^{\prime}$ are even, $K$ is odd. This theorem is proved in [4], the
proof is based on the classification of $P$-manifolds, $SP$-manifolds
and Lagrangian submanifolds given in this paper.

   Let us formulate also a theorem that permits us to give a description
of symmetry transformations in BV-formalism.

   Let us consider two operators $\tilde {\Delta}$ and $\Delta$ on a
supermanifold $M$ that determine different $SP$-structures, but the
same $P$-structure on $M$. Let us suppose that corresponding
volume elements $d\tilde {\mu}$ and $d\mu$ are connected by the
formula: $d\tilde{\mu}=e^{\sigma}d\mu$. Then

 a) $\Delta\exp({\sigma\over 2})=0$,

 b) If $S$ obeys the master equation $\Delta \exp S=0$ then $\tilde
{S}=S-{1\over 2} \sigma$ satisfies the master equation $\tilde {\Delta}
\exp \tilde {S}=0$.

 c)The action functional $\tilde {S}$ in the  $SP$-structure determined
by $\tilde {\Delta}$ describes the same physics as the action functional
$S$ in the  $SP$-structure determined by $\Delta$. In particular,

 \begin {equation}
 \int_L e^{\tilde {S}}d\tilde {\lambda} =\int _Le^Sd\lambda
 \end {equation}
 for every Lagrangian submanifold $L\subset M$.

 It follows immediately from this statement that every theory is
physically equivalent to the theory with an action functional $\tilde
{S}=0$. In this representation a symmetry is  simply an automorphism
of corresponding  $SP$-structure. For an arbitrary  $SP$-structure we
obtain:

  Every function $H$ satisfying $\Delta H+\{ H,S\}=0$ (every quantum
observable) determines a symmetry in the following sense. Neither the
the action functional $S$ nor the volume element $d\mu$ are invariant
with respect to an infinitesimal transformation with the Hamiltonian
$H$, however the new action functional

 $$\tilde {S}=S+\epsilon\{ H,S\}$$

 and the new volume element
 $$d\tilde {\mu}=d\mu(1+2\epsilon\Delta H)$$
 describe the same physics as the old action functional $S$ and the
old volume element $d\mu$.

    We will not discuss here interesting questions arising in the analysis
of the asymptotic behavior of $\int _L \exp (\hbar^{-1}S)d\lambda$ in
the $\lim \hbar \rightarrow 0$ (semiclassical approximation in
BV1formalism). Such an analysis was performed in [5]; the answer
was expressed in terms a generalization of Reidemeister torsion.

    \vskip .1in
  \centerline{{\bf References}}
  \vskip .1in
  1. Rosly, A.A., Schwarz, A.S.,Voronov,A.A.: Geometry of
Superconformal Manifolds, 1 and 2. Commun. Math. Phys. 119 (1988)
129, 120 (1989) 437.

  2. Rosly, A.A., Schwarz, A.S.: Supermoduli Spaces. Commun. Math.
Phys. 135  (1990) 91.

  3. Schwarz, A.S.: Symplectic, Contact and Superconformal
Geometry, Membranes and Strings. Trieste Supermembrane
Conference(1989) 332.

  4. Schwarz, A.S.: Geometry of Batalin-Vilkovisky Quantization.
Commun. Math. Phys. 155 (1993) 246.

  5. Schwarz, A.S.: Semiclassical Approximation in Batalin-Vilkovisky
Formalism. Commun. Math. Phys. 158 (1993) 373.

  6. Schwarz, A.S.: Symmetry  Transformations in Batalin-Vilkovisky
Formalism.  Lett. Math. Phys. (1994).

  7. Schwarz, A.S.: On the Definition of Superspace. Teor . Mat. Fiz. 60
(1984) 37.

  8. Friedan, D., Martinec, E. and Shenker, S.: Nucl. Phys. B 271
(1986) 93.

  9. Distler, J. and Nelson P.: Semirigid supergravity. Phys. Rev. Lett.
66 (1991) 1955.

  10. Getzler, E.; Batalin-Vilkovisky algebras and two-dimensional
topological field theories. Commun. Math. Phys. 159 (1994) 265

  11. Dijkgraaf, R., Verlinde, E., and Verlinde, H.: Topological strings in
$d<1$, Nucl. Phys. B352 (1991) 59.

\end {document}